# Vibrational Spectroscopy at Atomic Resolution with Electron Impact Scattering

Kartik Venkatraman[1], Barnaby D.A. Levin[1], Katia March[2], Peter Rez[3], and Peter A. Crozier[1*]

[1]School for Engineering of Matter, Transport and Energy, Arizona State University, Tempe, Arizona, USA

[2]Eyring Materials Center, Arizona State University, Tempe, Arizona, USA

[3]Department of Physics, Arizona State University, Tempe, Arizona, USA

*Corresponding Author: crozier@asu.edu

# Summary

Atomic vibrations control all thermally activated processes in materials including diffusion, heat transport, phase transformations, and surface chemistry. Recent developments in monochromated, aberration corrected scanning transmission electron microscopy (STEM) have enabled nanoscale probing of vibrational modes using a focused electron beam. However, to date, no experimental atomic resolution vibrational spectroscopy has been reported. Here we demonstrate atomic resolution by exploiting localized impact excitations of vibrational modes in materials. We show that the impact signal yields high spatial resolution in both covalent and ionic materials, and atomic resolution is available from both optical and acoustic vibrational modes. We achieve a spatial resolution of better than 2 Å which is an order of magnitude improvement compared to previous work. Our approach represents an important technical advance that can be used to provide new insights into the relationship between the thermal, elastic and kinetic properties of materials and atomic structural heterogeneities.

Atomic vibrations control all thermally activated processes in materials including ionic, atomic and electron diffusion, heat transport, phase transformations and surface chemical reactions. The jump frequency characterizing thermally activated processes is of great practical importance and is determined by the local phonon and molecular vibrational modes of the system. Atomic and molecular heterogeneities and defects such as vacancies, interstitials, dislocations and grain boundaries often regulate kinetic pathways and are associated with vibrational modes which are substantially different from bulk modes. High spatial resolution vibrational spectroscopy is required to probe these defect modes.

Recent developments in aberration corrected, monochromated, scanning transmission electron microscopy (STEM) have enabled nanoscale probing of vibrational modes via electron energy-loss spectroscopy (EELS)[1,2]. Nanoscale vibrational spectroscopy is already impacting a wide range of important scientific problems such as measurement of surface and bulk vibrational excitations in MgO nanocubes[3], probing hyperbolic phonon polaritons in nanoflakes of hBN[4], measuring temperature in nanometer-sized areas with 1°K precision[5,6] and determining phonon dispersion in nanoparticles[7]. The delocalized nature of certain vibrational signals allows damage-free nanoscale detection for a variety of organic and inorganic material-systems [8-11].

This progress has been impressive, however, to date there have been no experimental methods to spectroscopically probe individual vibrational modes in materials with atomic resolution. Theoretical treatments have explored the question of spatial resolution[12,13] with some treatments suggesting that atomic resolution vibrational EELS should be possible[14-16]. Here we demonstrate atomic resolution vibrational spectroscopy in STEM for signals predominantly excited by impact scattering. The resulting order of magnitude advance in spatial resolution will enable studies in which vibrational modes associated with single atomic columns and individual defects can be directly probed.

The vibrational spectrum associated with electron scattering has features in common with both photon and neutron spectroscopies. The electron interactions have been discussed in terms of dipole and impact scattering in low-energy high-resolution electron energy-loss spectroscopy (HREELS) by Ibach and others[17]. Dipole scattering is associated with the long-range Coulomb field which excites vibrational modes by polarizing the medium giving spectral features similar to those observed in infrared (IR) absorption spectroscopy[9,10]. The practical spatial resolution of dipole scattering is on the order of 20 – 200 Å and can be predicted with classical dielectric theories[18,19]. The electron may also undergo impact scattering, exciting vibrational modes that appear in neutron scattering but not IR spectroscopy; this includes acoustic modes in all materials, optical modes in non-ionic materials, and symmetric stretching and deformation modes in ionic materials[17]. Impact scattering is associated with short-range interactions and, in principle, should exhibit atomic resolution. A simple way to identify impact modes in EELS is to compare IR, neutron and electron vibrational spectra; impact modes will be present only in the neutron and electron spectra.

Two practical challenges must be resolved to perform atomic resolution vibrational EELS in a high performance monochromated STEM. First, a fundamental criterion for atomic resolution for any electron microscopy signal is that the experimental geometry must allow scattered electrons spanning a range of angles up to at least a Bragg angle to interfere[20]. For a STEM, this challenge can be addressed by appealing to the reciprocity principle first described by Cowley[21]. In STEM, an electron probe smaller than a (hkl) Miller plane spacing can only be formed if the incident electron beam has a convergence semi-angle of at least the Bragg angle. In the sample, momentum transfers will take place across such a convergent probe spanning at least one Bragg angle (or Brillouin zone boundary) ensuring that a high resolution signal will be collected by a spectrometer located on the optic axis of the microscope (see Fig. 1a). The convergent illumination condition required to create the small probe and achieve high spatial resolution necessarily means the EELS signal entering the spectrometer will consist of an integral over regions of momentum space spanning more than one Brillouin zone boundary. However, the experiments presented here also show that some momentum filtering does occur as the probe is scanned across a unit cell.

The second challenge is to suppress or avoid the dipole signal. For ionic materials, dipole signals, which often arise due to asymmetric stretching or deformation modes involving adjacent cations and anions, are many times stronger than impact signals[12]. Other researchers have approached this challenge using displaced spectrometer entrance apertures, achieving a spatial resolution of better than 20 Å[22]. However, in our approach, we perform vibrational spectroscopy at atomic resolution (<2 Å) using conventional on-axis EELS geometry. In non-ionic elemental semiconductors, such as Si, impact scattering predominates since any dipole signal will be a second order effect due to bond polarization by the passing fast electron. Since impact scattering dominates in Si, we show that it is possible to achieve atomic resolution by selecting either acoustic or optical phonons. For ionic materials, an alternative is selecting an impact scattering peak that lies at a different energy-loss than the dipole peaks. This signal can in principle yield atomic resolution and our experiments demonstrate that a localized impact signal in amorphous $SiO_2$ can be used to define the interface between amorphous $SiO_2$ and crystalline Si with 5 Å resolution.

Fig. 1b shows a typical experimental spectrum from Si, with a strong vibrational peak at ~60 meV and a secondary peak at ~45 meV. The characteristic vibrational signals sit on a rapidly falling background arising from both non-characteristic phonon losses and the tail of the zero-loss peak. This background was modelled and subtracted from the vibrational peaks (see Fig. 1b); technical details are given in Methods and examples are shown in Extended Data Fig. 1. The intensity and shape of the phonon spectra entering the spectrometer can be expressed as a product of a projected density of states and a position dependent transition probability. A theoretical description of electron scattering by phonons along with a schematic diagram showing the scattering geometry is given in Supplementary Information and Extended Data Fig. 2. Qualitative insights on the origin of spectral features can be ascertained from the phonon dispersion curves for Si, which have been determined using density functional theory[23,24] (reproduced in Fig. 1c), and experimentally[25]. The part of the phonon dispersion surface that is sampled during our experiment is associated with a cross-section of the Brillouin zone orthogonal to the beam direction as shown in Extended Data Fig. 3. Flat parts of the dispersion curves give rise to maxima in the phonon density of states[26] contributing to stronger spectral intensity. The higher energy peak in the spectrum (~60 meV) is associated with the upper transverse optical branches of the dispersion curves, which are shaded in blue in Fig. 1c, while the lower energy shoulder is associated with longitudinal acoustic and optical modes which are shaded red. The contributions from the high and low energy branches to the spectral intensity can be quantified with a simple peak fitting model. Two Gaussians, constrained to have widths between 10 and 30 meV, were fitted to background subtracted spectra, with peak positions constrained to lie in the ranges 40 – 50 meV and 55 – 63 meV, corresponding to the lower and higher energy bands described above. The two Gaussian model fits the experimental datapoints well (Fig. 1b).

Inelastic neutron scattering (Extended Data Fig. 4) shows similar features between 40 and 60 meV[27], and these are not observed with IR spectroscopy, confirming that the signals observed in EELS are associated with impact scattering and should be highly localized. To explore the localized nature of the vibrational EELS signals, the spectral intensity was investigated as a function of electron probe position by performing linescans across a Si unit cell. The images in Fig. 2a and b were recorded from Si in the [$\bar{1}10$] projection with the monochromator slit inserted to give an energy resolution of 10 meV with a spectrometer entrance aperture corresponding to a collection angle of 24 mrad. EELS linescans were performed along the [110] direction with a typical step size of about 0.2 Å. The total intensity entering the spectrometer passes through minima and maxima as the probe moves on and off the atomic columns due to interference effects associated with the phase contrast bright-field signal. To correct for this effect, the spectral intensity was normalized to the total intensity entering the spectrometer.

Fig. 2c shows the resulting vibrational spectra when the probe is positioned on and off the Si dumbbell columns and Fig. 2d shows the integrated intensity of the low energy and high energy peaks as the probe is moved across the unit cell. The spectral intensity of the 60 meV peak increases by almost 40%, when the probe is positioned on the column showing the impact signal is highly localized with Fig. 2d demonstrating a spatial resolution of better than 2 Å. This is at least an order of magnitude improvement in spatial resolution compared to previous results[3,19,22] clearly showing that atomic resolution vibrational spectroscopy can be accomplished in the forward scattering geometry using the impact signal. The signal from the lower energy peak is noisier but also shows atomic resolution.

When the collection semi-angle was reduced by a factor of two to 12 mrad, the intensity difference for the on and off column probe positions increased to 80% for the 60 meV line scan. Fig. 3 shows the spectra and integrated peak intensities as a function of position (linescan step size 0.6 Å) for this smaller collection angle. Interestingly, whilst the intensity of the higher energy signal still tracks with the HAADF signal peaking on the column, the maximum of the lower energy signal is offset from the column position. Furthermore, in contrast with the large collection angle data, the energies of both peaks change significantly as the probe moves between the columns. The lower energy peak increases in intensity to a maximum as the probe approaches the column, and its energy position shifts from ~ 40 to 50 meV. The higher energy peak shifts from 60 to 58 meV as the probe moves onto the column. This highlights the variation in intensity and shape of the spectra with less than 1 Å shifts in probe position.

For the small collection angle data, the lower energy peak shows an asymmetry in the intensity with respect to the atomic column position which also correlates with an asymmetry in the bright-field signal entering the spectrometer (see Extended Data Fig. 5a). Simulations of convergent beam patterns (Extended Data Fig. 5b-d) show that small (~1 mrad) misalignments between the incident cone and the cone defined by the spectrometer entrance aperture, as well as small (~1 mrad) tilts of the specimen can introduce an asymmetry to the intensity distribution in the bright-field signal recorded on either side of the atomic columns. Contributions to vibrational spectra from near the Brillouin zone boundaries may also be sensitive to small detector shifts or specimen tilts, which would explain the asymmetry we observe. This is an interesting effect and in future studies we will aim to exploit this sensitivity to tilts and shifts to investigate anisotropy in lattice vibrations with atomic resolution.

The low energy peak lies at ~ 50 meV when the electron probe is located close to the atomic column and is associated with the upper branches between the X and L points of the Brillouin zone (Fig. 1c). When the probe is located between the columns the energy shifts down to ~ 40 meV, which is associated with other points in the 2D ($\bar{1}10$) section of the Brillouin zone (see Extended Data Fig. 3). The change in energy suggests that the transition probability for launching phonons along different directions is strongly influenced by the probe position.

A form of momentum filtering also occurs for the higher energy optical peak and can be interpreted in terms of a simple classical picture where small impact parameter collisions are associated with high momentum transfer. When the probe is on the atomic column (small impact parameter), the peaks shift down to 56 – 58 meV corresponding to the higher momentum transfers associated with excitations at the Brillouin zone boundaries. When the probe is positioned between the columns (large impact parameter), the spectral peak appears at around 60 meV which is associated with relatively low momentum transfers in the first Brillouin zone (i.e. perhaps a third of the way between the Γ point and the boundary). Atomic resolution can arise from these low momentum transfer modes with pure elastic scattering followed by a normal phonon scattering process in the first Brillouin zone. Alternatively, it could arise via Umklapp scattering from neighboring Brillouin zones without the need for elastic scattering. Given that the sample thickness is on the order of the extinction distances (~ 50 nm) for Bragg beams, it is likely that the resulting signal with atomic resolution is a combination of both possibilities. Changes observed in the spectral shape with changing probe position can be explained by Umklapp scattering but not elastic scattering, thus showing that we have atomic resolution in the optical phonon mode.

High spatial resolution in vibrational EELS is also possible from ionic materials provided suitable impact peaks can be identified where the dipole contribution is very weak. $SiO_2$ is an oxide with mixed ionic-covalent bonding and the background-subtracted vibrational energy-loss spectrum (Fig. 4a) shows peaks at 58, 100 and 144 meV. The peak at 100 meV corresponds to a mixture of $SiO_4$ stretching and $SiO_4$ bending modes[28]. There is a weak dipole component from bond polarization effects, so it will not be prominent in IR absorption, but easily detectable by neutron measurements of impact scattering (see Extended Data Fig. 4)[28]. The spatial resolution of the $SiO_2$ peaks was explored by performing EELS linescans across a $SiO_2$/Si interface. The sharpness of the profile from the impact signal is controlled by $SiO_2$/Si interface abruptness. According to dielectric theory, the long-range electrostatic interaction and associated begrenzungs effect should cause the dipole signal to significantly drop at 20 – 30 Å from the interface[19]. An ADF image of the interface along with linescans of the 144 meV dipole and the 100 meV impact signals are shown in Fig. 4b and c. The EELS signals were normalized to the total spectral intensity to correct for elastic scattering and mass thickness effects. Whereas the 144 meV dipole signal shows delocalization of greater than 30 Å predicted by dielectric theory, the 100 meV impact signal shows a much sharper profile dropping over a distance of about 5 Å. This shows that high spatial resolution is not limited to elemental semiconductors but is possible in all materials (including amorphous materials) which possess peaks where impact scattering dominates over dipole scattering.

In conclusion, we have demonstrated atomic resolution vibrational spectroscopy by using a monochromated, aberration corrected STEM with a large convergence angle and a conventional on-axis spectrometer geometry. Our approach selects modes associated with impact scattering, which yield high spatial resolution for both ionic and covalent materials. We achieve a spatial resolution of better than 2 Å which is at least an order of magnitude improvement compared to previous work. In crystalline Si, with only covalent bonding, vibrational peaks corresponding to both optical and acoustic phonons were observed. The signal intensity is strongest with the probe positioned over atomic columns. We observed significant changes in spectral shape with sub-angstrom shifts in probe position for smaller collection angles, which is qualitatively explained by a momentum filtering effect. In amorphous $SiO_2$, with mixed ionic and covalent bonding, the 100 meV vibrational mode, which is dominated by impact scattering, yielded a spatial resolution of 5 Å. The method presented here represents an important technical advance that will allow the influence of local atomic structure on vibrational modes to be explored directly. In future, we will apply this method to investigate changes in the character of vibrational modes around atomic-scale structural heterogeneities such as point defects, dislocations, grain boundaries and surfaces. These changes in vibrational modes should yield fundamental new insights into important thermally activated processes in materials.

## Methods

Sample Preparation

To investigate vibrational spectra in Si, a conventional cross section Si sample in a [$\bar{1}$10] zone axis orientation was prepared by dimpling and ion milling. A second sample was prepared to investigate the Si/SiO$_2$ interface. The top surface of a Si wafer was subjected to thermal oxidation at 900°C to obtain a ~3 μm film of SiO$_2$ on the Si substrate. The oxidized wafer was then prepared for STEM EELS analysis by performing a lift-out procedure using a Ga-ion beam and an Omniprobe on a Nova 200 NanoLab (FEI) focused ion beam (FIB) combined with a scanning electron microscope (SEM). The thickness of the lift-out specimen measured using SEM approached ~100 nm near the edges and ~80 nm near the SiO$_2$/Si interface. The interface plane normal was parallel to the (001) crystallographic plane in Si while the zone axis was along the [110] crystallographic direction in Si.

Monochromated STEM-EELS Measurements

STEM-EELS analysis on all samples was performed using a NION HERMES UltraSTEM 100 aberration-corrected electron microscope equipped with a monochromator, operated at 60 kV. The probe convergence semi-angle was 28 mrad, and the corresponding collection semi-angles were 12 and 24 mrad. To record spectra, a dispersion of 1 or 2 meV per channel was used with the 12 mrad collection angle and 0.37 meV per channel was used with the 24 mrad collection angle. Aberration correction of the magnetic lenses up to the fifth order produced probes of ~0.12 nm diameter with beam currents of ~100 pA. During the monochromated experiment, the beam current was ~10 pA, and energy resolution was 15 meV for the 12 mrad collection angle and 10 meV for the 24 mrad collection angle. The monochromated probe size was estimated to be ~0.17 nm.

Background Fitting to Vibrational EELS Spectra

All acquired spectra were processed using the Gatan Microscopy Suite, the FIJI-Cornell Spectrum Imager[29], and custom written MATLAB code. The spectra were calibrated first by using cross correlation to align the center of the zero-loss peaks to 0 meV. All spectra were then normalized to the total intensity entering the spectrometer.

The characteristic vibrational peaks sit on a rapidly falling signal arising from both non-characteristic phonon losses and a tail on the zero-loss peak. The signal-to-background ratio is typically 10 – 20% making accurate background modelling and correction critical to reveal subtle differences in the characteristic peak shapes and intensities. Background subtraction in previous studies on vibrational EELS has typically involved fitting a power-law or polynomial function to the data using two fitting windows, one placed immediately before, and one placed immediately after the vibrational peak of interest. Earlier work on low-loss EELS has tested several functions that can be used to approximate the form of the zero-loss peak itself[30] finding the Pseudo-Voigt function (the sum of a Gaussian and a Lorentzian) to be particularly well suited.

In this work, we also used a two-window method for background fitting. We tested the exponential, power-law, and linear combination of power-laws (LCPL) functions because these have traditionally been used for background subtraction in EELS[29]. In addition, we tested the Pseudo-Voigt, and Pearson VII functions (essentially a Lorentzian raised to a power)[31] because these can be used to model spectral peaks, and so we hypothesized that they might be superior functions to fit to the tail of the zero-loss peak than exponential or power-law functions. When processing our experimental data, the most appropriate function for background subtraction was chosen based on three main criteria:

1) Minimization of a reduced $\chi^2$ value for the fit within the fitting windows (using the square root of the number of counts as a crude approximation of the noise).

2) Inspection of spectra to ensure that the intensities in signal channels are positive after background subtraction (negative intensity is unphysical and indicative of an incorrect background model).

3) Robustness of background subtraction when changing the position and width of the fitting windows (a significant drop in fit quality when repositioning windows is indicative of a poor background model).

For Silicon, we found the that Pseudo-Voigt function outperformed the other fitting models, because it resulted in fewer background subtracted spectra with regions of negative intensity, and was more robust to changes in the width and position of fitting windows.

For linescans across the $Si/SiO_2$ interface, we found that the Pearson VII function to be the most robust background model because it was the only function of those tested that did not leave a region of significant negative intensity after background subtraction between the vibrational peaks at ~58 and 144 meV when the probe was positioned at the interface. In this case, the fitting windows had to be spaced much further apart than they were for the spectra from Si, which may make accurate background fitting more challenging.

The MATALB codes used to fit backgrounds are uploaded as supplementary files.

Gaussian Fitting To Background Subtracted Si Vibrational Spectra

After background subtraction, we modeled the variation in the intensity of vibrational spectra from Si using a simple two Gaussian peak fitting model. Each Gaussian was constrained to have widths of up to a maximum of 40 meV, because this was the energy range over which the vibrational spectra of interest lay. A lower limit for Gaussian width was chosen based on the energy resolution of our measurements (10 meV when using a larger collection angle, and 15 meV when using a smaller collection angle). The peak positions were constrained to lie in the ranges 55 – 63 meV and 40 – 50 meV to model the higher and lower energy phonon dispersion branches in Si as indicated in Fig. 1c.

The MATLAB code used to fit Gaussians to the spectra are uploaded as supplementary files.

Calculating the phonon dispersion surfaces

The 2D dispersion surfaces were calculated with a combination of Phonopy and VASP using a 4 x 4 x 4 supercell based on the primitive Si unit cell. The VASP calculations were converged to $1.0 \times 10^{-8}$ eV with the PAW potentials. Phonon frequencies were calculated on a grid of 307 points spanning one quadrant of the $\bar{1}10$ 2D Brillouin Zone[32-34].

Data Availability

The MATLAB codes used to analyze the data are included as supplementary files. The raw experimental spectra will be made freely available by depositing them in a suitable repository prior to publication.

**References:**

1. Krivanek, O. L. *et al.* Vibrational spectroscopy in the electron microscope. *Nature* **514**, 209-212. (2014).
2. Miyata, T. *et al.* Measurement of vibrational spectrum of liquid using monochromated scanning transmission electron microscopy-electron energy loss spectroscopy. *Microscopy* **63**, 377-382, doi:10.1093/jmicro/dfu023 (2014).
3. Lagos, M. J., Trugler, A., Hohenester, U. & Batson, P. E. Mapping vibrational surface and bulk modes in a single nanocube. *Nature* **543**, 529-+, doi:10.1038/nature21699 (2017).
4. Govyadinov, A. A. *et al.* Probing low-energy hyperbolic polaritons in van der Waals crystals with an electron microscope. *Nature Communications* **8**, 10, doi:10.1038/s41467-017-00056-y (2017).
5. Idrobo, J. C. *et al.* Temperature Measurement by a Nanoscale Electron Probe Using Energy Gain and Loss Spectroscopy. *Phys. Rev. Lett.* **120**, 4, doi:10.1103/PhysRevLett.120.095901 (2018).
6. Lagos, M. J. & Batson, P. E. Thermometry with Subnanometer Resolution in the Electron Microscope Using the Principle of Detailed Balancing. *Nano Lett.* **18**, 4556-4563, doi:10.1021/acs.nanolett.8b01791 (2018).
7. Hage, F. S. *et al.* Nanoscale momentum-resolved vibrational spectroscopy. *Sci. Adv.* **4**, 6, doi:10.1126/sciadv.aar7495 (2018).
8. Jokisaari, J. R. *et al.* Vibrational Spectroscopy of Water with High Spatial Resolution. *Adv. Mater.* **30**, 6, doi:10.1002/adma.201802702 (2018).
9. Rez, P. *et al.* Damage-free vibrational spectroscopy of biological materials in the electron microscope. *Nature Communications* **7**, 7, doi:10.1038/ncomms10945 (2016).
10. Haiber, D. & Crozier, P. A. Nanoscale Probing of Local Hydrogen Heterogeneity in Disordered Carbon Nitrides with Vibrational EELS. *ACS Nano* **12**, 5463–5472, doi:DOI 10.1021/acsnano.8b00884 (2018).
11. Crozier, P. A., Aoki, T. & Liu, Q. Detection of water and its derivatives on individual nanoparticles using vibrational electron energy-loss spectroscopy. *Ultramicroscopy* **169**, 30-36, doi:http://dx.doi.org/10.1016/j.ultramic.2016.06.008 (2016).
12. Rez, P. Is Localized Infrared Spectroscopy Now Possible in the Electron Microscope? *Microscopy and Microanalysis* **20**, 671-677, doi:doi:10.1017/S1431927614000129 (2014).
13. Egerton, R. F. Prospects for Vibrational-Mode EELS with High Spatial Resolution. *Microscopy and Microanalysis* **20**, 658-663, doi:doi:10.1017/S1431927613014013 (2014).
14. Forbes, B. D. & Allen, L. J. Modeling energy-loss spectra due to phonon excitation. *Phys. Rev. B* **94**, doi:10.1103/PhysRevB.94.014110 (2016).
15. Dwyer, C. Localization of high-energy electron scattering from atomic vibrations. *Phys. Rev. B* **89**, 5, doi:10.1103/PhysRevB.89.054103 (2014).
16. Rez, P. Does Phonon Scattering Give High-Resolution Images *Ultramicroscopy* **52**, 260-266, doi:10.1016/0304-3991(93)90034-u (1993).
17. Ibach, H. & Mills, D. L. *Electron energy loss spectroscopy and surface vibrations* (Academic Press, 1982).

18    Crozier, P. A. Vibrational and valence aloof beam EELS: A potential tool for nondestructive characterization of nanoparticle surfaces. *Ultramicroscopy* **180**, 104-114, doi:https://doi.org/10.1016/j.ultramic.2017.03.011 (2017).
19    Venkatraman, K., Rez, P., March, K. & Crozier, P. The influence of surfaces and interfaces on high spatial resolution vibrational EELS from SiO2. *Microscopy* **67**, 14 -23, doi:DOI 10.1093/jmicro/dfy003 (2018).
20    Spence, J. C. H. *High Resolution Electron Microscopy, 3rd Edition*. Vol. 60 (Oxford Science Publications, 2003).
21    Cowley, J. M. Image Contrast in a Transmission Scanning Electron Microscope *Applied Physics Letters* **15**, 58-&, doi:10.1063/1.1652901 (1969).
22    Dwyer, C. *et al.* Electron-Beam Mapping of Vibrational Modes with Nanometer Spatial Resolution. *Phys. Rev. Lett.* **117**, doi:10.1103/PhysRevLett.117.256101 (2016).
23    Jain, A. *et al.* Commentary: The Materials Project: A materials genome approach to accelerating materials innovation. *Apl Materials* **1**, doi:10.1063/1.4812323 (2013).
24    Ong, S. P. *et al.* Python Materials Genomics (pymatgen): A robust, open-source python library for materials analysis. *Computational Materials Science* **68**, 314-319, doi:https://doi.org/10.1016/j.commatsci.2012.10.028 (2013).
25    Dolling, G. Lattice vibrations in crystals with the diamond structure. (Atomic Energy of Canada Ltd., Chalk River, Ont., 1962).
26    Ashcroft, N. W. & Mermin, N. D. *Solid State Physics*.  (W.B. Saunders Company, 1976).
27    Kulda, J., Strauch, D., Pavone, P. & Ishii, Y. Inelastic-neutron-scattering study of phonon eigenvectors and frequencies in Si. *Phys. Rev. B* **50**, 13347 (1994).
28    Haworth, R., Mountjoy, G., Corno, M., Ugliengo, P. & Newport, R. J. Probing vibrational modes in silica glass using inelastic neutron scattering with mass contrast. *Phys. Rev. B* **81**, 060301, doi:10.1103/PhysRevB.81.060301 (2010).
29    Cueva, P., Hovden, R., Mundy, J. A., Xin, H. L. & Muller, D. A. Data Processing for Atomic Resolution Electron Energy Loss Spectroscopy. *Microscopy and Microanalysis* **18**, 667–675 (2012).
30    Zhu, J. T., Crozier, P. A., Ercius, P. & Anderson, J. R. Derivation of Optical Properties of Carbonaceous Aerosols by Monochromated Electron Energy-Loss Spectroscopy. *Microscopy and Microanalysis* **20**, 748-759, doi:10.1017/s143192761400049x (2014).
31    Hall, M., Veeraraghavan, V., Rubin, H. & Winchell, P. The approximation of symmetric X-ray peaks by Pearson type VII distributions. *J. Appl. Crystallogr.* **10**, 66-68 (1977).
32    Kresse, G. & Furthmüller, J. Efficient iterative schemes for ab initio total-energy calculations using a plane-wave basis set. *Physical Review B* **54**, 11169-11186 (1996).
33    Kresse, G. & Joubert, D. From ultrasoft pseudopotentials to the projector augmented-wave method. *Physical Review B* **59**, 1758 (1999).
34    Togo, A. & Tanaka, I. First principles phonon calculations in materials science. *Scripta Materialia* **108**, 1-5 (2015).
35    Arai, M. *et al.* High resolution S (Q, E) measurement on g-SiO2. *Physica B: Condensed Matter* **180**, 779-781 (1992).
36    Dolling, G. & Cowley, R. The thermodynamic and optical properties of germanium, silicon, diamond and gallium arsenide. *Proceedings of the Physical Society* **88**, 463 (1966).

37  Lehmann, A., Schumann, L. & Hübner, K. Optical Phonons in Amorphous Silicon Oxides. I. Calculation of the Density of States and Interpretation of LO-TO To Splittings of Amorphous SiO2. *Physica status solidi (b)* **117**, 689-698 (1983).
38  Koch, K. *QSTEM: Quantitative TEM/STEM Simulations*, <https://www.physics.hu-berlin.de/en/sem/software/software_qstem> (2018).
39  Amali, A. & Rez, P. Theory of lattice resolution in high-angle annular dark-field images. *Microscopy and Microanalysis* **3**, 28-46 (1997).

**End notes**

Supplementary Materials

Supplementary Information

Extended Data Fig. 1 – 5

References (29 – 39)

Acknowledgments

Financial support for K.V., B.D.A.L., P.R. and P.A.C. was provided by US NSF CHE-1508667 and for B.D.A.L. and P.A.C. by Department of Energy grant DE-SC0004954. We also acknowledge the use of (S)TEM at John M. Cowley Center for High Resolution Electron Microscopy in the Eyring Materials Center at Arizona State University. P.A.C. would like to acknowledge stimulating discussions on atomic resolution vibrational spectroscopy with Prof. Les Allen. We acknowledge assistance from Dr. Arunima Singh in the use of Phonopy.

Author Contributions

K.V. prepared samples, K.V. and K.M. acquired all experimental vibrational EELS data, B.D.A.L. developed software for spectral processing, K.V. and B.D.A.L. analyzed EELS results. B.D.A.L performed simulations of CBED patterns, K.V. performed dielectric theory simulations and P.R developed and interpreted phonon models. P.A.C. and P.R. initiated the project and were involved in extensive discussion on the interpretation of the results. All authors were active in writing the manuscript.

Competing Interests

The authors declare no competing financial interests.

**Figures and Figure legends**

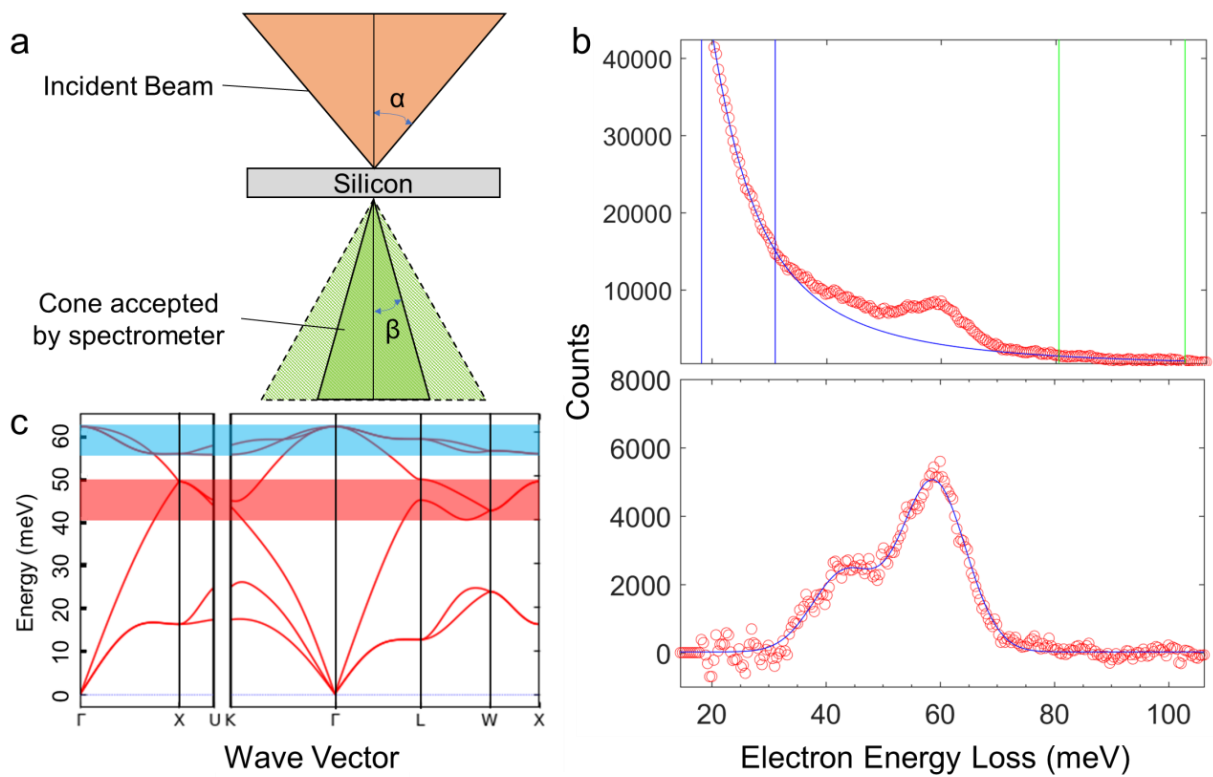

**Fig. 1. Acquisition geometry and vibrational spectra from silicon. a** Schematic diagram showing the STEM EELS acquisition geometry used in experiment with α = 28 mrad and β = 12 or 24 mrad. **b** The upper panel shows a typical raw vibrational energy-loss spectrum from Si acquired with β = 24 mrad (red circles). Blue curve shows the background subtraction model employed. The vertical lines (blue and green) show the windows used to fit the background. The lower panel shows the same spectrum after background subtraction. Here, the blue curve shows the two Gaussian peak fitting model employed, which shows good agreement with the data. **c** Dispersion surfaces in for Si[23,24] – blue and red shaded areas corresponds to peaks in the density of states on the upper branches from ~55 – 62 meV (optical) and on a lower branch from ~41 – 48 meV (optical and acoustic).

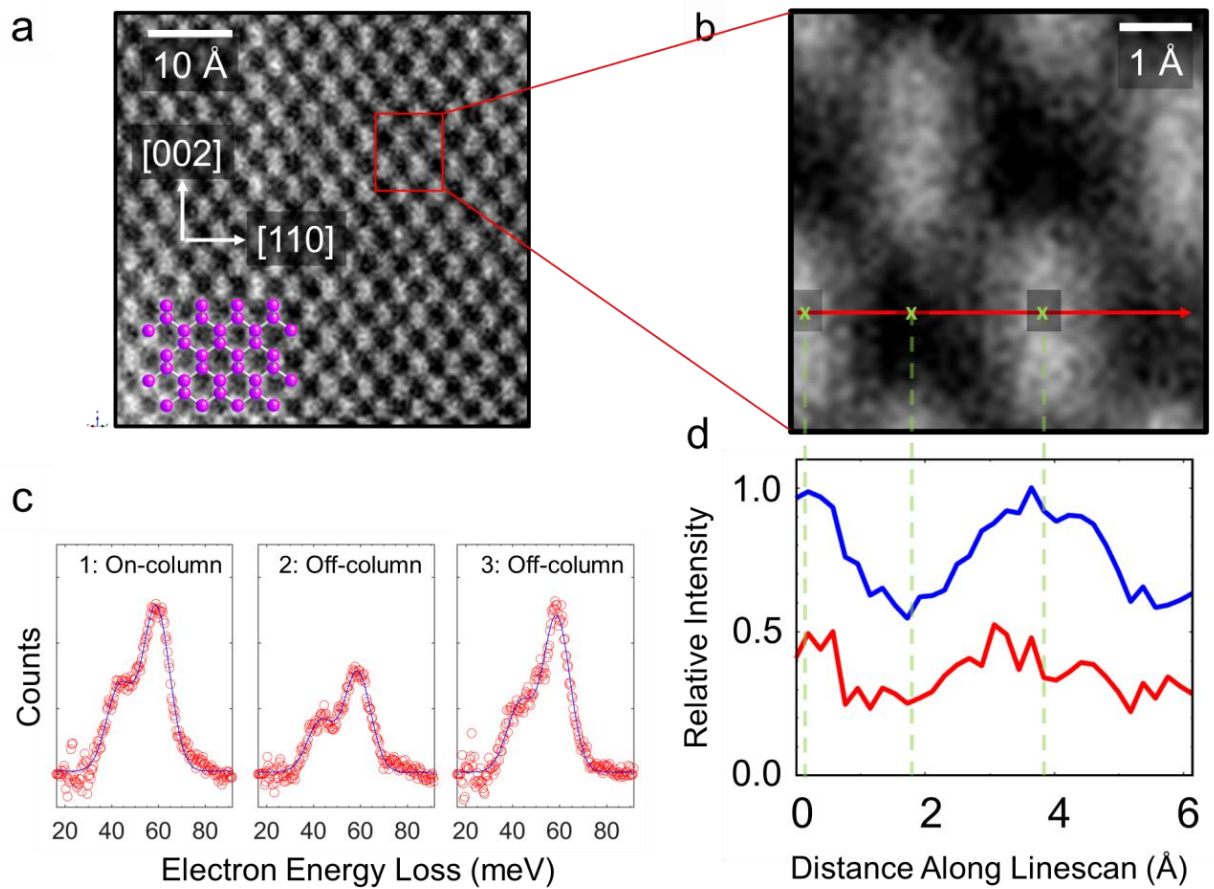

**Fig. 2. Background subtracted spectra and linescan for large collection angle data. a** Atomic resolution annular dark field (ADF) image with electron beam monochromation of ~10 meV from Si in [$\bar{1}10$] projection with light contrast corresponding to Si dumbbell column pairs. A model of Si in the [$\bar{1}10$] projection overlaid on the ADF image shows the position of atomic columns that form dumbbells. **b** Magnified image of area indicated by red box in a. The arrow indicates position and direction of EELS linescan acquisition. **c** Individual spectra from different positions along linescan shown by the green x-symbols in b. As in Fig. 1, the red circles are experimental data points, and the blue line is the result of the Gaussian fit. **d** The variation in intensity for higher (blue) and lower (red) energy phonons over the linescan indicated in b. The profiles are spatially aligned with the ADF image in b, as indicated by dashed green guidelines. An increase in phonon intensity can be observed around the dumbbell columns.

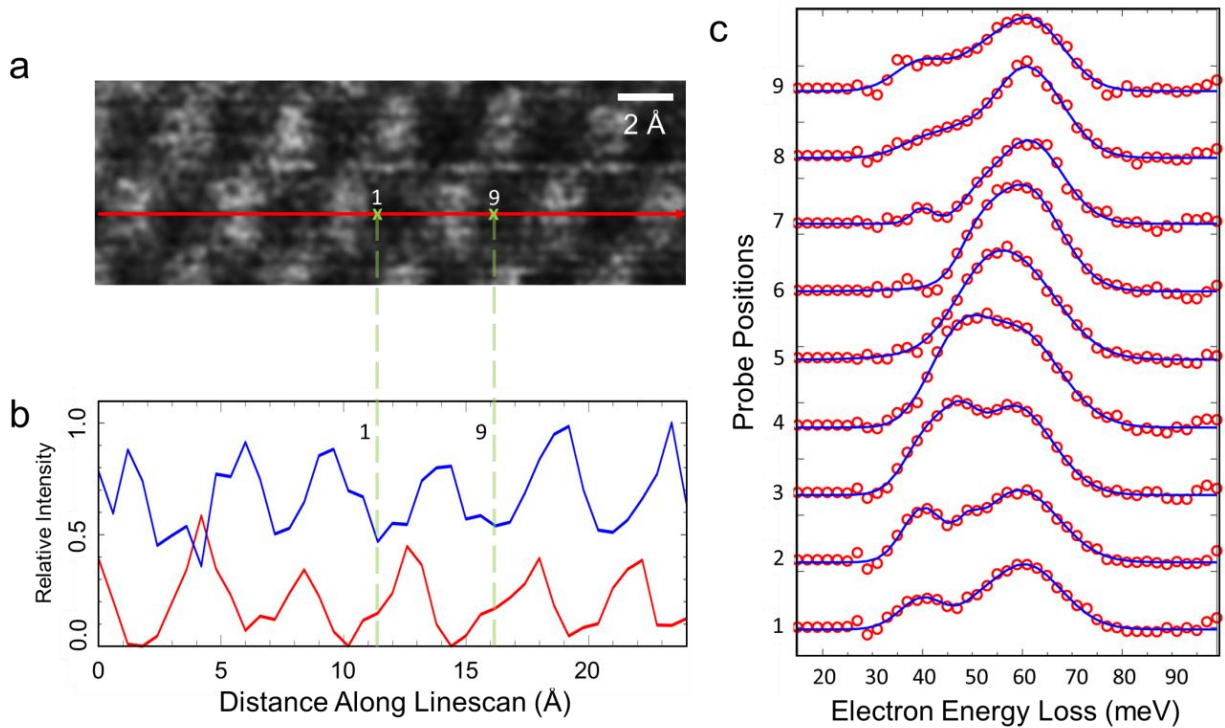

**Fig. 3. Background subtracted spectra and linescans for small collection angle data. a** ADF image of Si dumbbells with arrow indicating position and direction of EELS linescan acquisition. **b** The variation in intensity for higher (blue) and lower (red) energy phonons over the linescan indicated in a. A strong increase in higher energy phonon intensity can be observed around the dumbbell columns. The maxima of the lower energy phonon intensity are offset relative to the dumbbell columns. **c** Raw spectra at 9 individual probe positions all separated by 0.6 Å along the linescan between the labels 1 and 9 shown in a and b. It also shows the two Gaussian fits to the spectra at all probe positions. This highlights the variation in intensity and shape of the spectra with less than 1Å shifts in probe position.

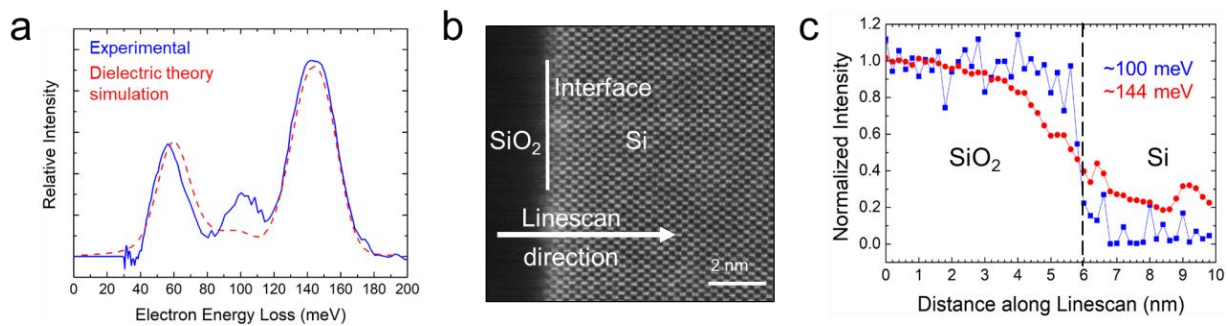

**Fig. 4. High resolution vibrational spectroscopy in SiO$_2$**. **a** Experimental energy-loss spectrum in SiO$_2$ far from interface (solid blue) and dielectric theory simulation of the spectrum (dashed red). The peak at ~100 meV does not appear strongly in the dielectric simulation, indicating that it is predominantly excited by impact scattering. **b** Atomic resolution ADF image of the SiO$_2$/Si interface showing linescan direction across the interface. **c** Normalized signal profiles across the interface – 100 meV (red) and 144 meV (blue).

**Extended Data**

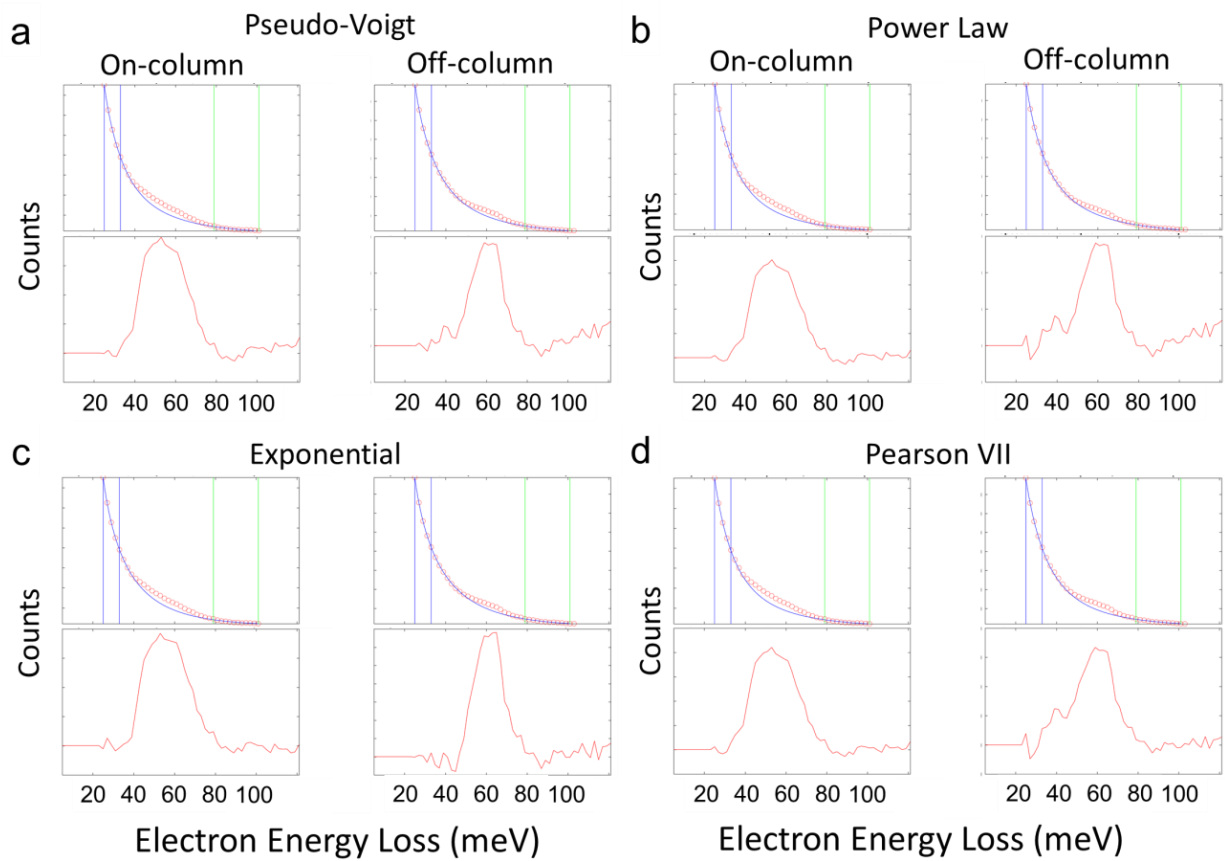

**Extended Data Fig. 1. Background fitting models.** Example two-window background fits and background subtracted spectra for Si with the probe in an on-column probe position (on the leading edge of a column immediately before the signal from the lower energy phonon cuts off) and an off-column position. The background models used are **a** a Pseudo-Voigt (Gaussian plus Lorentzian), **b** a power-law, **c** an exponential, and **d** a Pearson VII function (essentially a Lorentzian raised to a power). A linear combination of power-laws (LCPL) model was also tested but is not shown because it performed very similarly to the power-law model. The width and position of the fitting windows are the same in each spectrum and are indicated by the vertical blue and green lines. Each of the background models yields background subtracted spectra with features that are qualitatively extremely similar. The exponential, power-law, and Pearson VII fits are judged to be inferior to the Pseudo-Voigt fit because of the negative intensity between ~20 and 40 meV energy loss in the on-column spectra.

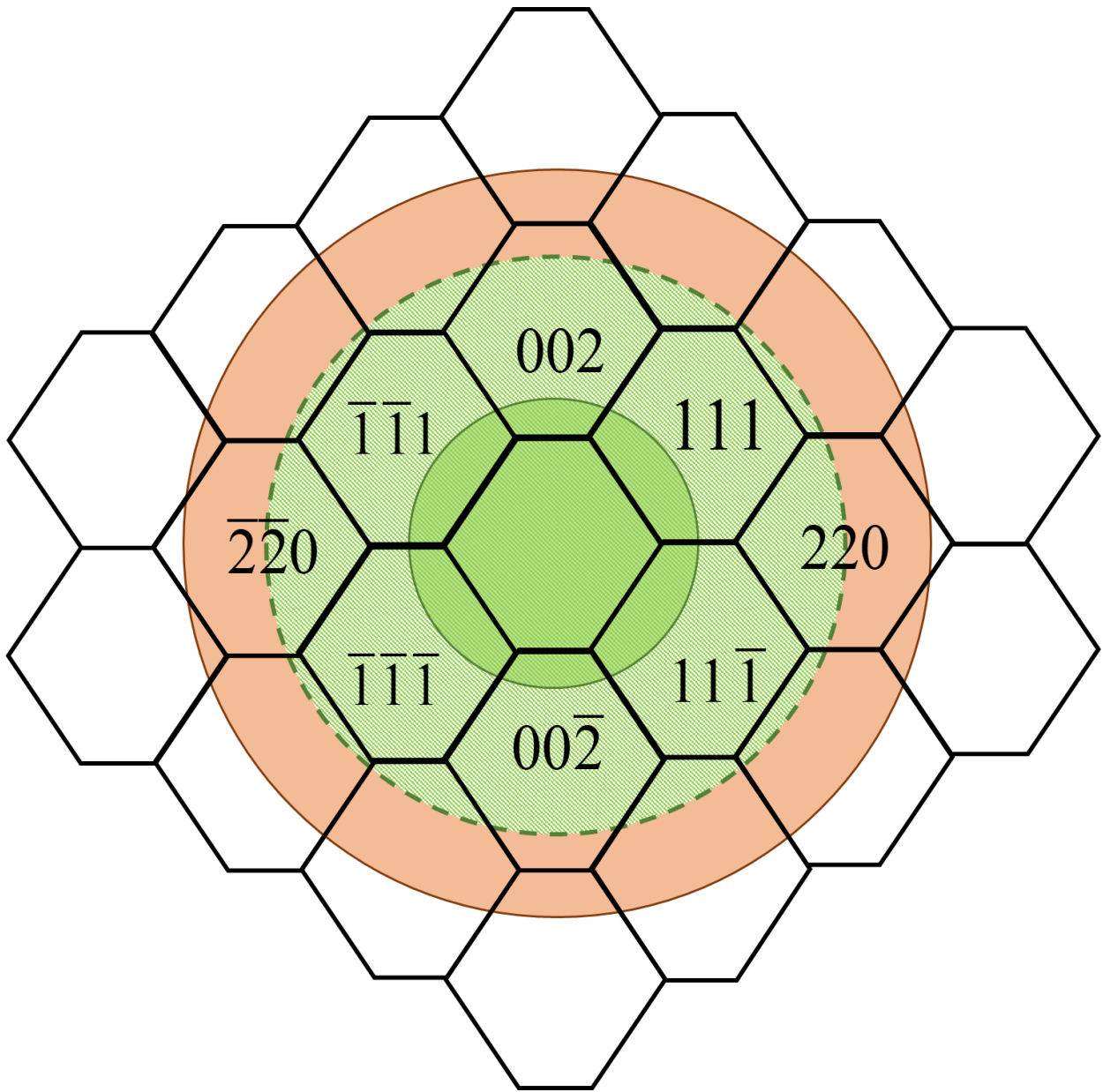

**Extended Data Fig. 2** Schematic diagram in reciprocal space showing incident beam convergence (orange circle) of 28 mrad and spectrometer acceptance ranges of 24 and 12 mrad (dashed and solid green circles respectively) relative to the Brillouin zones in Si covered by the incident beam.

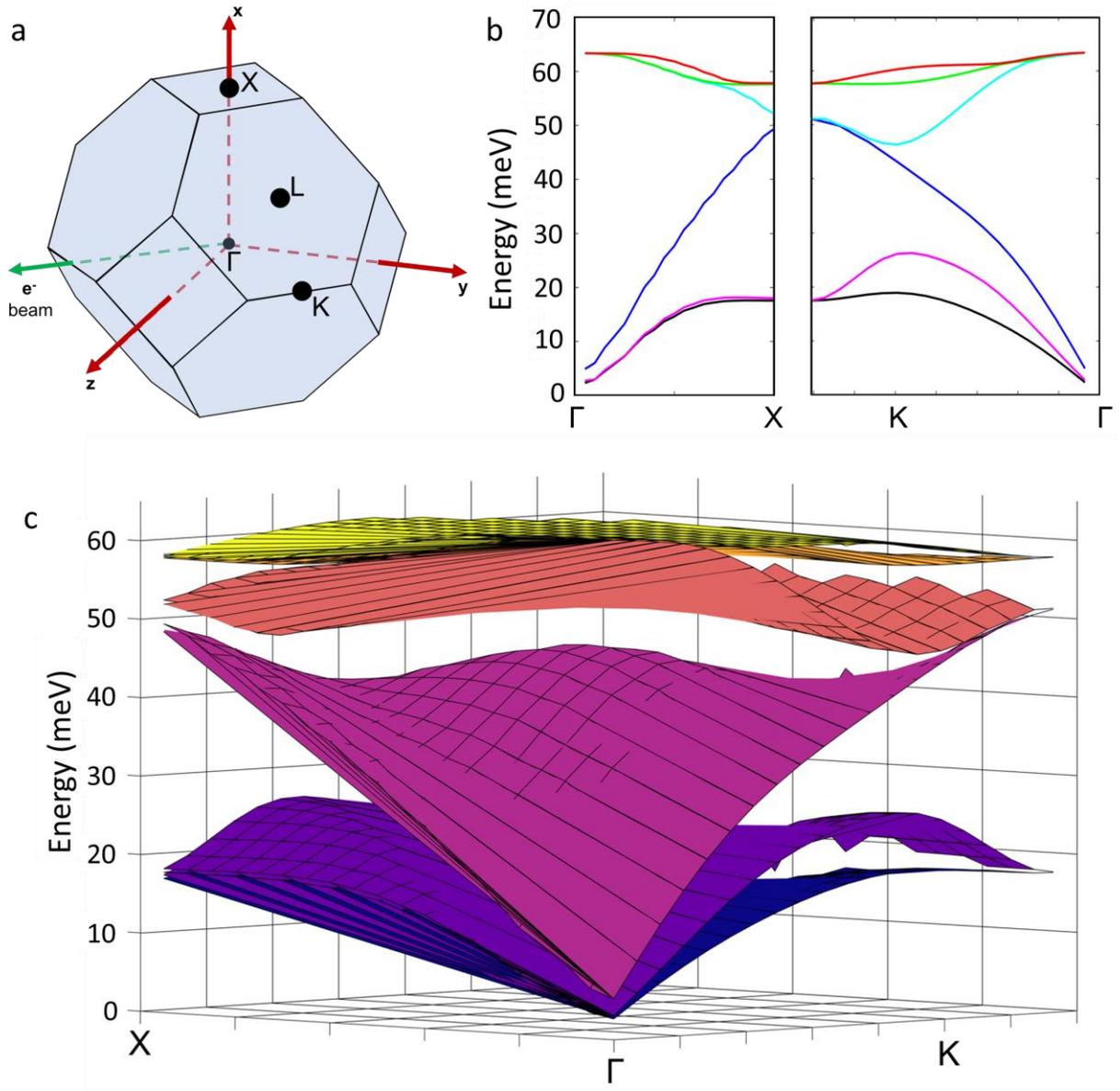

**Extended Data Fig. 3**. **a** Schematic diagram of the Brillouin Zone of a face-centered cubic unit cell. In our experiments, the electron beam propagates along a direction symmetrically equivalent to [$\bar{1}$10]. The part of the phonon dispersion surface that is sampled during our experiment is associated with a cross-section of the Brillouin zone orthogonal to the beam direction, which intersects the Γ, X, K, and L points indicated on the diagram. **b** 1D plots of sections of the phonon dispersion curves between Γ and X, and Γ and K, calculated using Phonopy. These show very similar features to the dispersion curves derived from the full 3-dimensional Brillouin zone (Fig. 1c). **c** A plot of the 2D dispersion surfaces associated with the cross-section of the Brillouin zone orthogonal to the beam direction, calculated using Phonopy. The higher energy phonon signal measured in our experiments is associated with the upper dispersion surfaces around 60 meV. The lower energy phonon signal measured in our experiments is associated with maxima, minima, and saddle points in the dispersion surfaces between 40 and 50 meV.

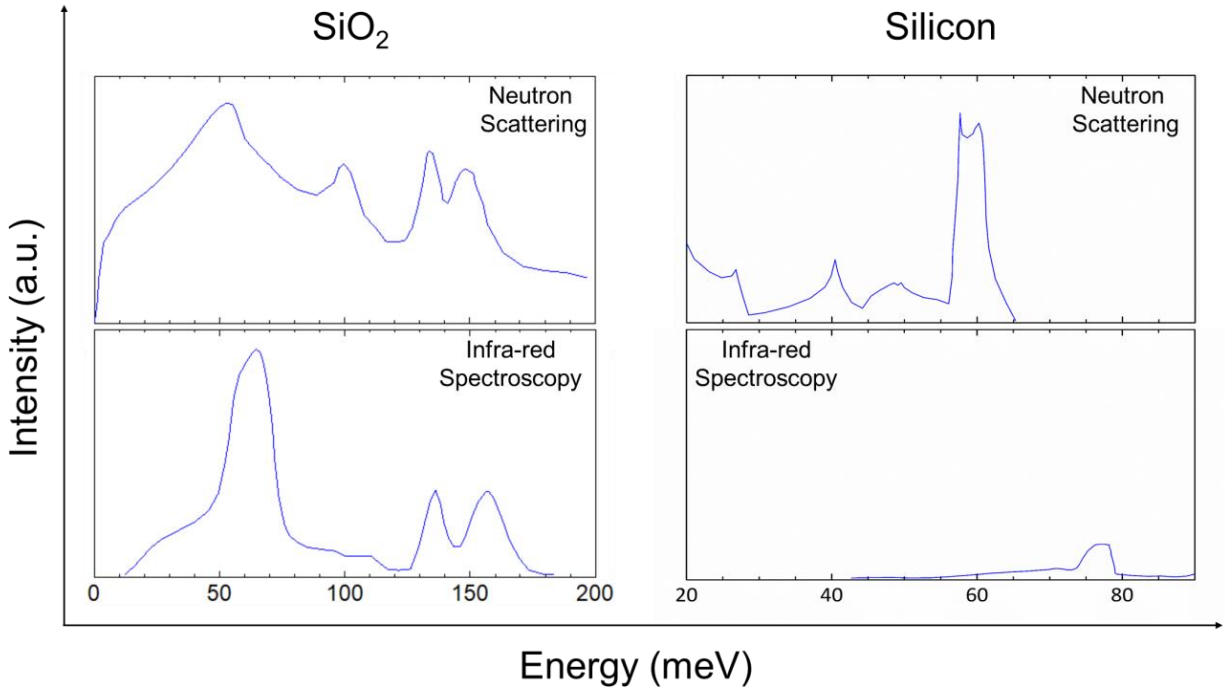

**Extended Data Fig. 4. Neutron and IR data.** To facilitate the interpretation of the vibrational EELS data we examined inelastic neutron scattering data for $SiO_2$[35] and $Si$[36] and IR absorption spectroscopy from $SiO_2$[37] and $Si$[36]. For Si, the relevant peaks are at approximately 40, 50 and 60 meV and are consistent with the dispersion curves shown in Fig. 1c. These peaks all also present with different relative intensities in the atomic resolution EELS. The IR peak at 75 meV is extremely weak and is not observed in the EELS spectra (indeed, simulations suggest that even with no background from the zero-loss peak, the 75 meV peak would be obscured by Cerenkov radiation). For $SiO_2$, the neutron scattering shows peaks at 55, 100, 138 and 150 meV. The IR shows three peaks at similar energies but the peak at 100 meV is absent since it is primarily associated with impact scattering. The EELS shows peaks like those in the neutron scattering although we do not resolve the separate peaks at 138 and 150 meV.

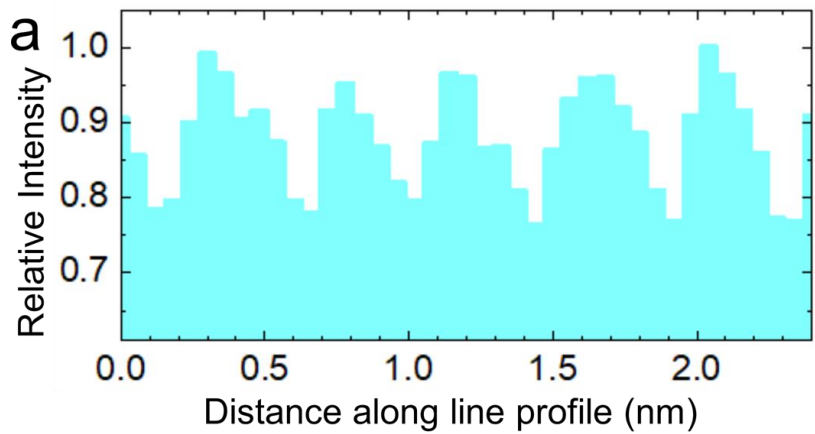

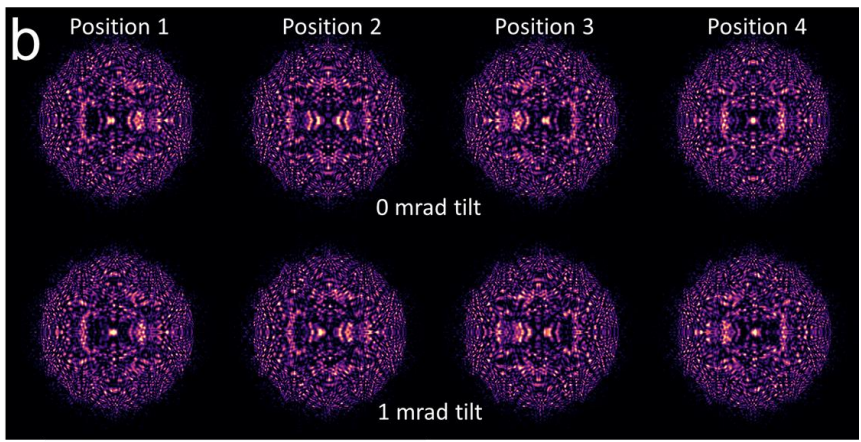

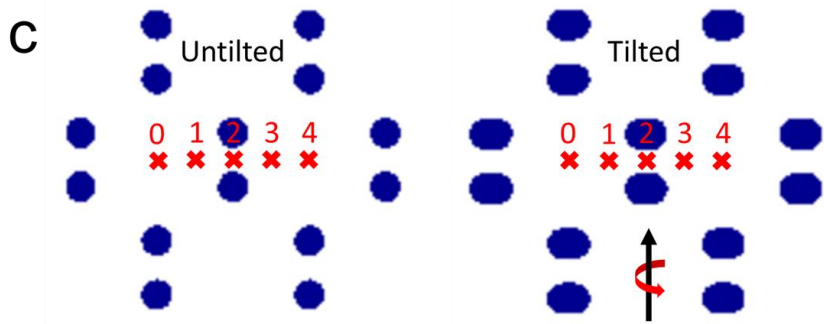

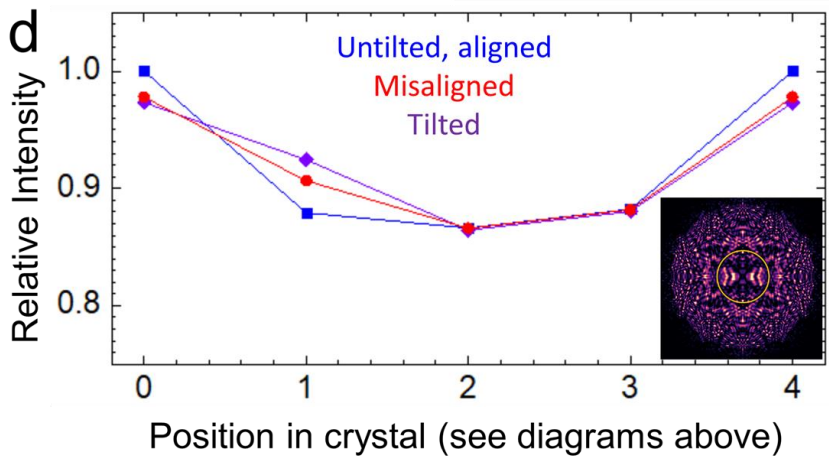

**Extended Data Fig. 5. Bright field signal: Experiment and Simulations. a** Total intensity entering the spectrometer passes through minima and maxima as the probe moves on and off the atomic columns due to destructive and constructive interference associated with the phase contrast bright field signal. The bright-field signal profile also shows an asymmetry which correlates with the asymmetry shown for the phonon intensities of Fig. 3b. **b** We performed energy filtered convergent beam diffraction simulations with a probe size of 2 Å and a convergence semi-angle of 28 mrad. The simulations were performed with Q-STEM[38] exactly on and at a tilt of 1 mrad off the (110) zone axis. Aberrations were chosen to fix the probe size at 2 Å for a 60 kV beam voltage. As we used a monochromated probe for our experiments, the energy spread of the beam is simulated as 0.0 eV. Thermal diffuse scattering was not included. **c** Diagrams indicating the probe positions corresponding to those of the simulations in (b) at 4 different points along the (110) linescans. **d** Simulated bright field signal entering the spectrometer vs probe position. This was determined by integrating the CBED patterns over a circular area corresponding to a collection semi-angle of 12 mrad (example area indicated by yellow circle on CBED pattern in inset). Positions 0 and 4 are symmetrically identical. The simulated on axis BF profile (blue line) shows a drop in intensity of about 13 % when the probe is on the Si dumbbell column which is reasonably close with the experimental conditions, where we have about 20% intensity change. Tilting the crystal by 1 mrad (purple line, tilt axis illustrated by arrows in (c)) introduces very little change to the intensity when the probe is on the column (Position 2) but causes an asymmetric change in the intensity on either side of the column (Positions 1 and 3). A similar effect is also seen due to a 1 mrad shift in the spectrometer entrance aperture position (red line). A combination of a small tilt and a small detector misalignment may be responsible for the asymmetry observed in the phonon linescan. Experimentally, it is very challenging to ensure that such small sample tilts and detector misalignments are not present.

**Supplementary Information**

Theoretical Model

The specimen thicknesses where measurements were performed was on the order of 50 nm, which is similar to the extinction distance for Si (110) (~ 55 nm). There is therefore dynamical elastic diffraction both before and after the phonon scattering. In many ways the process is similar to HAADF where lattice resolution is achieved by coherent interference in the incident probe, followed by transfer to a detector by multiphonon (thermal diffuse) scattering[39]. In this case the transfer is to the range of angles and energies selected by the spectrometer, by both acoustic and optic phonons. The phonon intensity when the probe is at position $r_p$ is an integral over contributions from slices of thickness $dz$ at depth $z$ in a specimen of thickness $t$

$$I(r_p) = \sum_{L,L',g,g',h,h'} \iiint Q_{0L}(q',t-z) Q^*_{0L'}(q',t-z)$$
$$\times H(q'+L-g-q) H^*(q'+L'-g'-q)$$
$$\times P_{gh}(q,z) P^*_{g'h'}(q,z) A(q+h) A(q+h') exp[-i(h-h')r_p] dq' dq\, dz$$

(1)

where $q'$ represents the wavevector in the Brillouin Zone (BZ) characterizing the final state accepted by the spectrometer, $q$ the wave vector in the BZ representing the initial state (See **Fig. 1**), $q'$-$q$ is the phonon wavevector $L,L'$, $g,g',h,h'$ are reciprocal lattice vectors, $A(q+h)$ is an aperture function defining the incident probe, $P_{gh}(q,z)$ represents dynamical propagation of the incident electron wave before the phonon scattering which is represented by $H(q'+L-g-q')$ and $Q_{0l}(q',t-z)$ represents the dynamical scattering of the phonon scattered electrons.

The scattering operator for phonons is

$$H(q+g) = (q+g).e \left(\frac{\hbar}{2M\omega(q)}\right)^{\frac{1}{2}} f_{el}(q+g)(1+exp(iq.R))$$

(2)

where $f_{el}(q+g)$ is the electron scattering factor, $R$ is the position of the 2nd atom in the primitive cell, $M$ is the mass of the atom, $\omega(q)$ is the frequency and $e$ is the polarization direction. We have assumed that energies that can be resolved by our spectrometer are greater than $k_BT$ (25 meV) and only phonon creation processes need be considered. In a Bloch wave picture the dynamical diffraction propagation matrices can be expressed as

$$P_{gh}(q,z) = \sum_j C_g^j exp(ik_j z) C_h^{*j} \quad (3)$$

where $k_j$ are the eignenvalues and $C_g^j$ are the eigenvectors. Alternatively the propagation matrix can be calculated by the multislice method.

In the simplified theory we'll neglect the dynamical elastic scattering before and after the phonon scattering, and assume that just 3 beams contribute to the image –g,0,g (In Si these could be (111) beams). Neglecting cross aperture interference we have

D$q = q'- q$

is the net phonon momentum transfer.

The simplified intensity $I(r_p)$ is

$$I(r_p) = \iint \left(\frac{2\hbar}{\omega M}\right) \cos^2(\Delta q R_p) \begin{bmatrix} f(\Delta q + g)f(\Delta q)((\Delta \boldsymbol{q} + \boldsymbol{g}).\boldsymbol{e})(\Delta \boldsymbol{q}.\boldsymbol{e})exp(-i\boldsymbol{g}.\boldsymbol{r_p}) \\ +f(\Delta q - g)f(\Delta q)((\Delta \boldsymbol{q} - \boldsymbol{g}).\boldsymbol{e})(\Delta \boldsymbol{q}.\boldsymbol{e})exp(i\boldsymbol{g}.\boldsymbol{r_p}) \\ (\Delta \boldsymbol{q}.\boldsymbol{e})^2 f^2(\Delta q) \end{bmatrix} \Delta q$$

(4)

which can be rewritten as an expression inversely proportional to the phonon energy, $E$, and a density of phonon states.

$$I(r_p) = \iint \left(\frac{2\hbar^2}{M}\right) \cos^2(\Delta q R_p) \begin{bmatrix} f(\Delta q + g)f(\Delta q)((\Delta \boldsymbol{q} + \boldsymbol{g}).\boldsymbol{e})(\Delta \boldsymbol{q}.\boldsymbol{e})exp(-i\boldsymbol{g}.\boldsymbol{r_p}) \\ +f(\Delta q - g)f(\Delta q)((\Delta \boldsymbol{q} - \boldsymbol{g}).\boldsymbol{e})(\Delta \boldsymbol{q}.\boldsymbol{e})exp(i\boldsymbol{g}.\boldsymbol{r_p}) \\ (\Delta \boldsymbol{q}.\boldsymbol{e})^2 f^2(\Delta q) \end{bmatrix} \frac{1}{E} \frac{dE}{|\nabla E(\Delta q)|}$$

(5)

Note that if there is a slight tilt such that the 1st two terms in the square brackets aren't equal then there will be a change in the peak phonon intensity with respect to $r_p$.